\documentclass{aa}  

\usepackage{graphicx}
\usepackage{txfonts}
\usepackage{natbib}
\usepackage{hyperref}

\makeatletter
\renewcommand*\aa@pageof{, page \thepage{} of \pageref*{LastPage}}
\makeatother
\begin{document}

        \title{Subarcsecond view on the high-redshift blazar GB\,1508+5714 by the International LOFAR Telescope}
        \titlerunning{Subarcsecond view on GB\,1508+5714 by the ILT}

\author{A. Kappes\inst{1}, P.\,R. Burd\inst{1}, M. Kadler\inst{1}, G. Ghisellini\inst{2}, E. Bonnassieux\inst{1}, M. Perucho\inst{9,10}, M. Brüggen\inst{3}, C.\,C. Cheung\inst{4}, \\B. Ciardi\inst{5}, \mbox{E. Gallo\inst{6},} \mbox{F. Haardt\inst{2,7,8},} L.~K. Morabito\inst{12,13}, T. Sbarrato\inst{2,11}, A. Drabent\inst{14}, J. Harwood\inst{15}, N.~Jackson\inst{16}, \\J. Moldon\inst{17}
          }
\authorrunning{A. Kappes et. al}

   \institute{Institut für Theoretische Physik und Astrophysik, Universität W\"urzburg,\\
              Emil-Fischer-Straße 31, 97074 W\"urzburg, Germany
              \email{alexander.kappes@uni-wuerzburg.de}
              \and
              INAF – Osservatorio Astronomico di Brera, Via Bianchi 46, I–23807 Merate, Italy
              \and
              Hamburger Sternwarte, Universit\"at Hamburg, Gojenbergsweg 112, 21029 Hamburg, Germany
              \and
              Space Science Division, Naval Research Laboratory, Washington, DC 20375, USA
              \and
              Max-Planck-Institut f\"ur Astrophysik, Karl-Schwarzschild-Straße 1, 85748 Garching b. M\"unchen, Germany
              \and
              Department of Astronomy, University of Michigan, 1085 S University, Ann Arbor, MI 48109, USA
              \and
              INFN – Sezione Milano–Bicocca, Piazza della Scienza 3, 20126 Milano, Italy
              \and
              Dipartimento di Scienza e Alta Tecnologia, Università dell’Insubria, Via Valleggio 11, 22100 Como, Italy
              \and
              Departament d'Astronomia i Astrof\'isica, Universitat de Val\`encia, C/ Dr. Moliner, 50, E-46100 Burjassot, Val\`encia, Spain 
              \and
              Observatori Astron\`omic, Universitat de Val\`encia, C/ Catedr\`atic Beltr\'an 2, E-46091 Paterna , Val\`encia, Spain
              \and
              Dipartimento di Fisica G. Occhialini, Univ. Milano–Bicocca, P.za della Scienza 3, I–20126 Milano, Italy
              \and
              Centre for Extragalactic Astronomy, Department of Physics, University of Durham, South Road, Durham DH1 3LE, UK
              \and
              Institute for Computational Cosmology, Department of Physics, University of Durham, South Road, Durham DH1 3LE, UK
              \and
              Th\"uringer Landessternwarte, Sternwarte 5, D-07778 Tautenburg, Germany
              \and
              Centre for Astrophysics Research, University of Hertfordshire, College Lane, Hatfield AL10 9AB, UK
              \and
              Department of Physics and Astronomy, School of Natural Sciences, University of Manchester, Manchester M13 9PL, UK
              \and
              Instituto de Astrof\'isica de Andaluc\'ia (IAA, CSIC), Glorieta de las Astronom\'ia, s/n, E-18008 Granada, Spain
             }

   \date{Received TBD; accepted TBD}

  \abstract
{Studies of the most distant active galactic nuclei (AGNs) allow us to test our current understanding of the physics present in radio-jetted AGNs across a range of environments, and to probe their interactions with these environments. The decrease in apparent luminosity with distance is the primary difficulty to overcome in the study of these distant AGNs, which  requires highly sensitive instruments.
}
{Our goal is to employ new long wavelength radio data to better parametrise the broad-band spectral energy distribution (SED) of GB\,1508+5714, a high-redshift ($z$=4.30) AGN. Its high redshift, high intrinsic luminosity, and classification as a blazar allow us to test emission models that consider the efficient cooling of jet electrons via inverse Compton losses in interactions with the dense cosmic microwave background (CMB) photon field at high redshifts. A significant detection of this effect in GB\,1508+5714 may partly explain the apparent sparsity of high-redshift radio galaxies in wide-field surveys;  detections of this kind are  only becoming possible with the current generation of Square Kilometre Array (SKA) precursors.
}
{We used the LOw-Frequency ARray (LOFAR) to image the long wavelength radio emission around the high-redshift blazar GB\,1508+5714 on arcsecond scales at frequencies between 128\,MHz and 160\,MHz. This allowed us to compare the spatially resolved structure with higher frequency observations, and to construct spectral index maps to study the spectral properties of the different components.
}
{The LOFAR image shows a compact unresolved core and two resolved emission regions around 2 arcsec to the east and to the west of the radio core. We find structure consistent with previous Very Large Array (VLA) observations, as well as a previously unreported emission region to the east. The region in the west shows a spectral index of $-1.2^{+0.4}_{-0.2}$, while the region in the east indicates a spectral index of $\lesssim-1.1$. The radio core features a flat spectral index of $0.02\pm0.01$.
}
{We interpret the arcsecond-scale radio structure of GB\,1508+5714 as a FR II-like radio galaxy at a small viewing angle, and the western component as the region containing the approaching jet’s terminal hot spot, while the eastern diffuse component near the core can be interpreted as the counter-hot spot region. Our SED modelling shows that a scenario featuring significant quenching effects caused by interaction with the CMB provides a good description of the data, and notably explains the suppressed radio emission.
}
\keywords{
        galaxies: active --
        galaxies: jets --
        galaxies: individual: GB\,1508+5714 --
        radio continuum: galaxies --
        techniques: high angular resolution --
        techniques: interferometric
}

\maketitle

%
\section{Introduction}
\label{sec:intro}
The remarkably bright nature of radio-jetted active galactic nuclei (AGNs) allows us to observe them at extreme redshifts, and thus use their properties as an observational tracer of cosmological principles \citep{wang2021}. Blazars, in particular, benefit from increased apparent luminosities due to relativistic boosting effects \citep{cohen2007}, and can thus be detected over a wide range of the electromagnetic spectrum, although at the cost of being observationally compact objects with small angles between their jets and the line of sight.
Conversely, AGNs with radio jets that are at a larger angle to the line of sight are harder to detect with increasing distance. Because they can be observed over such a wide range of redshifts, they provide a unique insight into cosmology, galaxy evolution, and the evolution of AGNs \citep{dunlop1990,georgakakis2017}. In addition to probing the universe by observing targets at different redshifts, one should also account for differences in their evolutionary stage, the environment they are embedded in at that time, and their interactions with that environment. 
Many radio surveys have already been performed to investigate radio-loud AGN populations \citep[e.g.][]{becker1995, condon1998, cohen2007a, intema2017} and most find self-consistent relative number ratios of radio galaxies and blazars up to a redshift of $\sim 3$ \citep{volonteri2011}. Beyond this redshift, matters are complicated by uncertainties about the density evolution and the build-up of high black hole masses in the early universe  \citep{blundell1999,shankar2008}. However, there seems to be a consensus that there is a relative lack of higher redshift radio galaxies even when accounting for evolutionary effects and detection limits \citep[e.g.][and references therein]{Wu2017,Hodges-Kluck2021}. The reason for the deficit is still not fully understood. The currently  favoured explanation was suggested by \cite{ghisellini2015}: interaction of the extended radio emission with the cosmic microwave background (CMB) could efficiently quench the brightness of the extended radio lobes.  \cite{morabito2018} find evidence to support this model based on a comparison of simulations and observational data.
In this scenario the CMB energy density dominates over the magnetic energy density at very high redshifts, so that the jet electrons interact with the CMB photons by inverse Compton (IC) scattering to cool, while the synchrotron radiation is suppressed. Although quenched, the steep-spectrum isotropic radiation of the extended structures could nevertheless be detected by telescopes operating in the  long wavelength radio regime, which can test and guide the theoretical models.
\cite{ghisellini2015} proposed a number of suitable blazars  to spotlight this issue, and published expected radio fluxes for different model parameters. 
The International LOFAR Telescope (ILT) \citep{vanhaarlem2013} offers the resolution, sensitivity, and observing wavelengths necessary to detect the extended emission from these blazars and to address the questions associated with its possible suppression. 
For this study we processed and  analysed one ILT observation of \object{GB\,1508+5714} at $z=4.30$ \citep{hook1995}, which is one of the most distant quasars with a detected X-ray jet \citep{yuan2003, siemiginowska2003}.
Throughout the paper we use the following cosmological parameters: $H_0=71$\,km$\cdot$s$^{-1}\cdot$Mpc$^{-1}$, $\Omega_\text{m}=0.27$, and $\Omega_\Lambda=0.73$, hence a luminosity distance of $39.8$\,Gpc and a conversion scale where 1" is about 6.9\,kpc for the given source.

\section{Observation and data reduction}
\label{sec:data_observation}
\begin{figure}[ht]
        \centering
        \includegraphics[width=\linewidth]{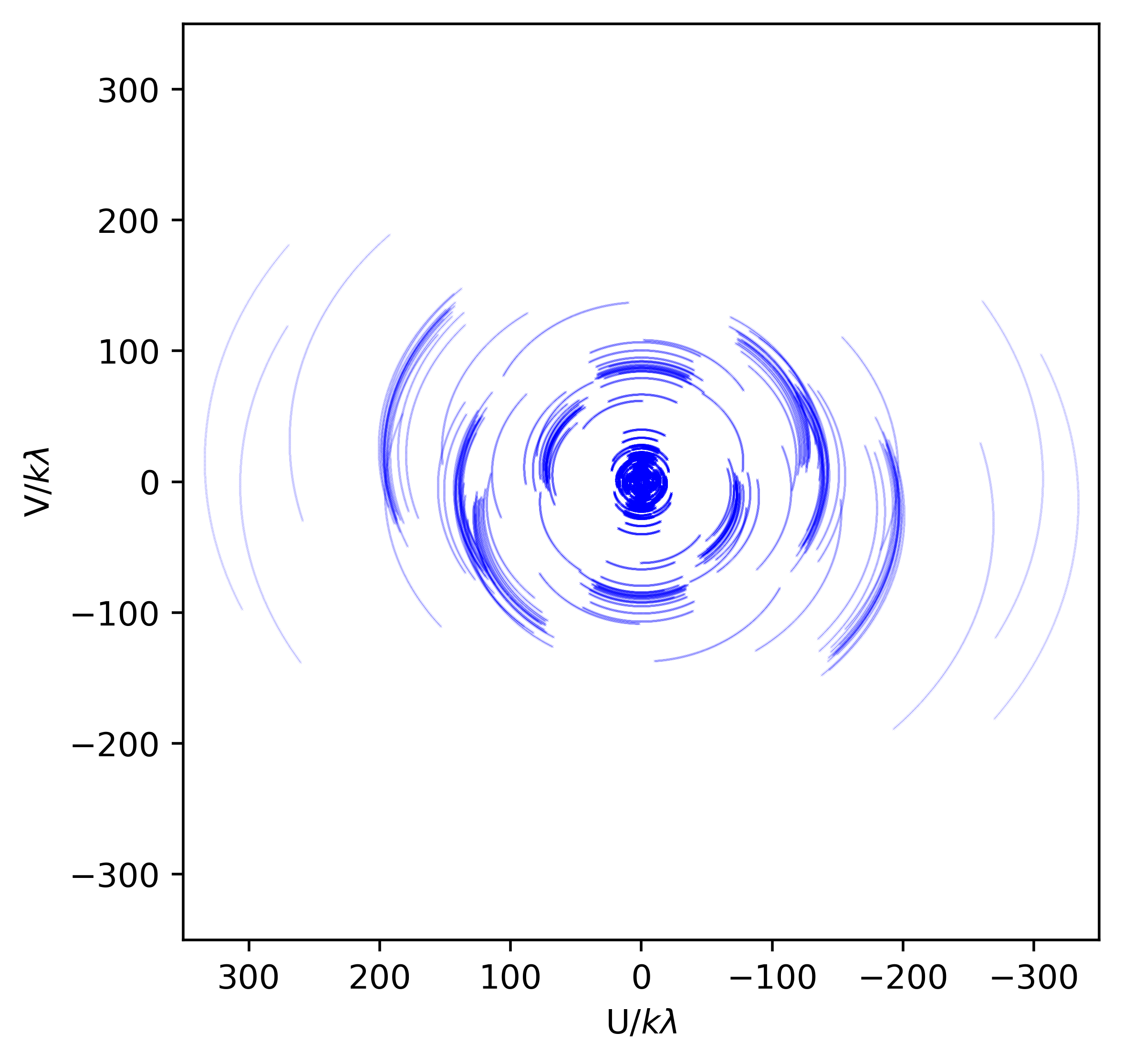}
        \caption{$(uv)$ coverage for our International LOFAR Telescope data set at 144\,MHz with the target GB\,1508+5714 located in the phase centre observed on 15 June  2015 for 4\,hours.
        }
        \label{fig:uvplane}
\end{figure}

We observed \object{GB\,1508+5714} on 15 June  2015 with the High Band Antenna (HBA) array of the international LOFAR telescope in dual outer mode\footnote{Each core station has two HBA tile-fields, which are treated as individual stations in this mode.}, with the target positioned in the phase  centre. Nine international stations\footnote{DE601HBA, DE602HBA, DE603HBA, DE604HBA, DE605HBA, DE609HBA, FR606HBA, SE607HBA, UK608HBA} participated in this observation.
The observation time was 4 hours covering 110\,MHz to 190\,MHz with applied time averaging of 16\,s and a frequency channel width of $12.2\,$kHz. 3C196 was used as a flux density calibrator, with a ten-minute calibration observation prior to the target observation.
After retrieving the data from the  Long Term Archive (LTA), it was processed with \textsc{prefactor}\footnote{\href{https://github.com/lofar-astron/factor}{Github repository https://github.com/lofar-astron/factor}} \citep{degasperin2019} version 3.0 to obtain calibration solutions for all stations using the calibrator, for polarisation alignment, clock, bandpass, and rotation measures. Phase solutions were found for the core and remote stations in the target field. This is a necessary step before continuing with the LOFAR-very long baseline interferometry (VLBI) pipeline\footnote{\href{https://github.com/lmorabit/lofar-vlbi}{Github repository https://github.com/lmorabit/lofar-vlbi}}, as described in \citet{morabito2021}, which uses these solutions to start the phase calibration of the international stations. After calibration, the final frequency coverage ranged between between 128\,MHz and 160\,MHz in four sub-bands with 8\,MHz bandwidth. This reduction in bandwidth was due to the presence of strong radio-frequency interference (RFI) in the data. The final full-band image has a central frequency of 144\,MHz. The final time-averaging was 16\,s. 
All core stations were combined to one virtual super station.
The resultant reduction in data size allows for a simplified data handling and reduces I/O  significantly \cite[for more in-depth information regarding the whole LOFAR-VLBI pipeline, see][]{morabito2021}. The remote stations RS503HBA, RS407HBA, and  RS406HBA, and  the international stations FR606HBA and  SE607HBA were also removed from the dataset because of the poor data quality of visibilities associated with these stations. This could be due to  problems at the station level at the time of observation or to issues with the model of 3C196 used in the initial calibration. Diagnostic plots show good ionospheric conditions overall, which suggests that the issues were driven by the quality of the calibrator models at the time of reduction.
If this is indeed the case, we expect the problem to disappear in the future when better models are   available. 
The final (uv) coverage is shown in Fig. \ref{fig:uvplane}.

\begin{figure}[ht]
        \centering
        \includegraphics[width=\linewidth]{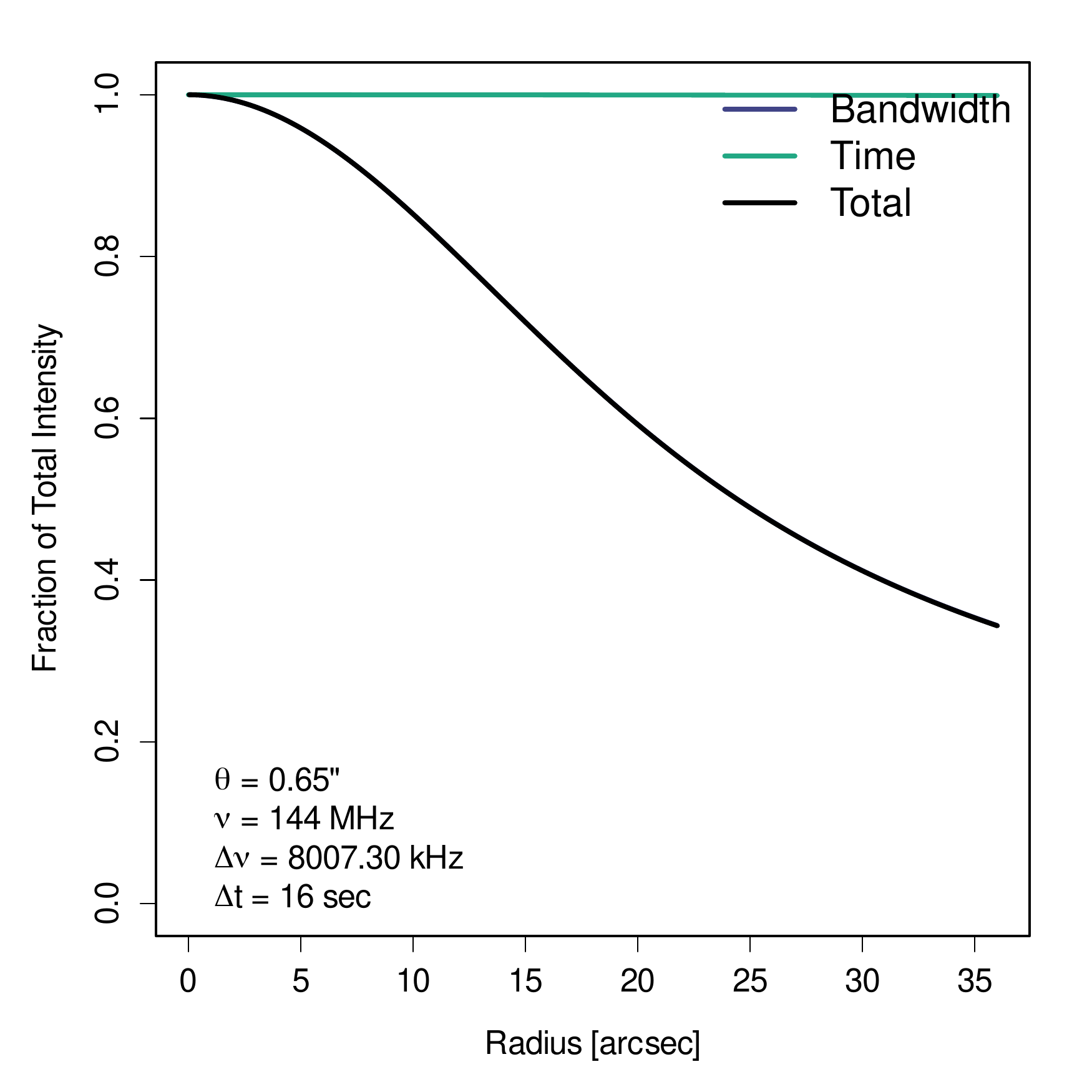}
        \caption{Bandwidth and time smearing losses affecting the intensity for a point source in the pointing centre as a function of radius. Calculated after \cite{bridle1999}. The blue line indicating the bandwidth smearing is hidden by the black line indicating the total smearing.
        }
        \label{fig:smearing}
\end{figure}

The imaging process was performed manually, with the addition of further self-calibration, using \textsc{difmap} \citep{shepherd1997}. To test the robustness of our results, three parties performed the imaging process independently without prior knowledge of the expected structure; they also  began the self-calibration with different {\bf starting} models to ensure that the final result was independent of the initial model. The reality of structure in the eastern direction was verified by not including this structure in the model supplied to the self-calibration step, and was verified  that it persisted despite this exclusion. Time and bandwidth smearing do not have a great impact on the final image because our target is located at the phase centre. We calculate the loss of intensity as a function of distance from the phase centre using the equations 18--43 for time smearing and 18--24 for bandwidth smearing from \cite{bridle1999}, with the final extinction curve shown in Fig. \ref{fig:smearing}. We clearly see that smearing remains below 5\% at a radial distance of about 6'', which corresponds to the total extent of the source. This effect can thus be considered  negligible in our subsequent analysis.

\begin{figure}[ht]
        \centering
        \includegraphics[width=\linewidth]{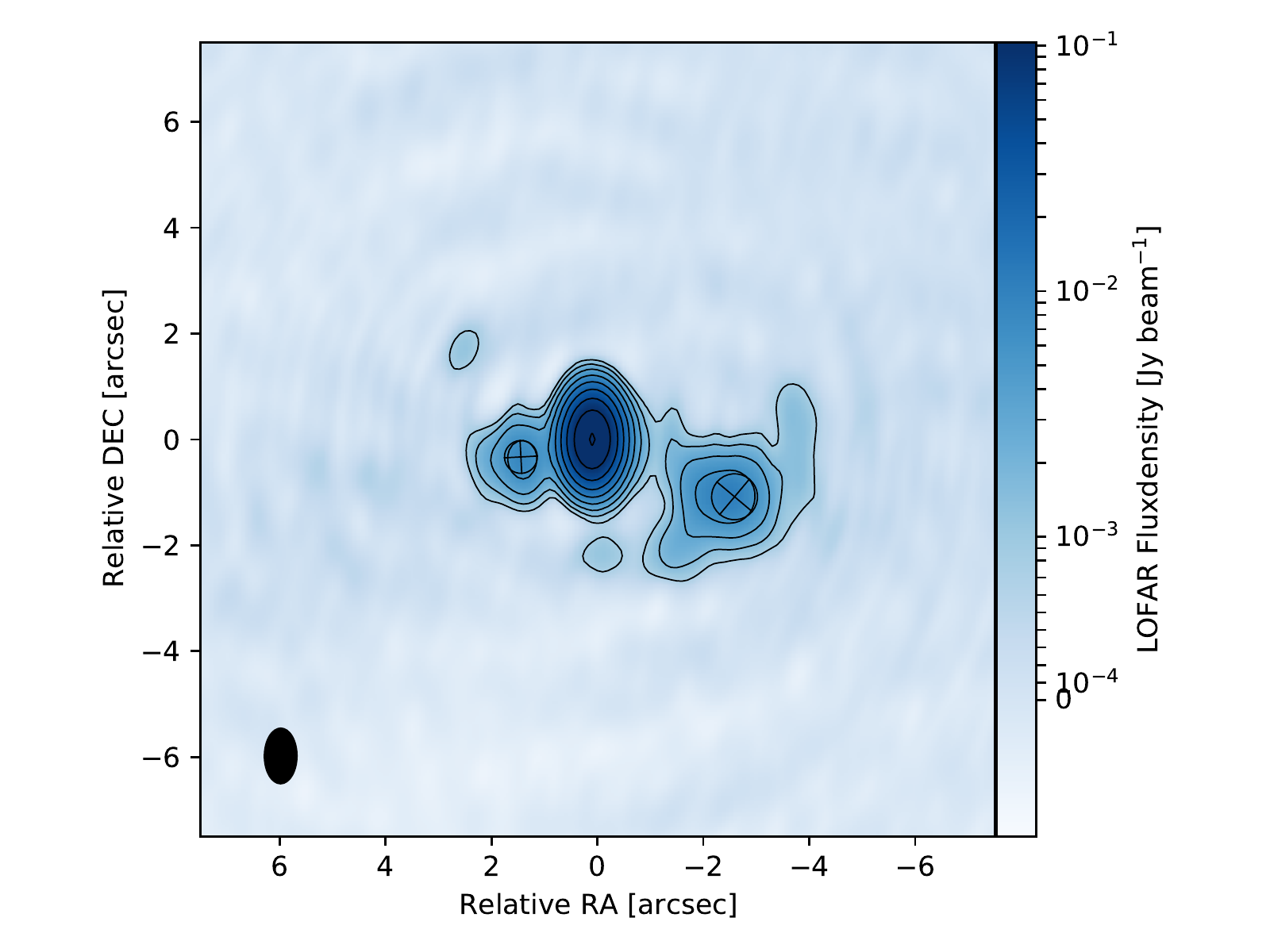}
        \caption{LOFAR image of GB\,1508+5714 at 144\,MHz. The source brightness is  colour-coded by flux density, with the contour lines overlaid. The crossed circles indicate the modelled Gaussian components as hot spots. The contour levels are drawn at ($-1, 1, 2, 4, 8,$ ... ) times $0.78$\,mJy$\,$beam$^{-1}$. The RMS is 0.13\,mJy$\,$beam$^{-1}$. The restoring beam size (shown in the bottom left) is $1.08'' \times 0.643''$ with a position angle of $-0.27^\circ$. 
        }
        \label{fig:lofarobs}
\end{figure}

\section{Results}
\label{sec:results}
\begin{figure}[ht]
        \centering
        \includegraphics[width=\linewidth]{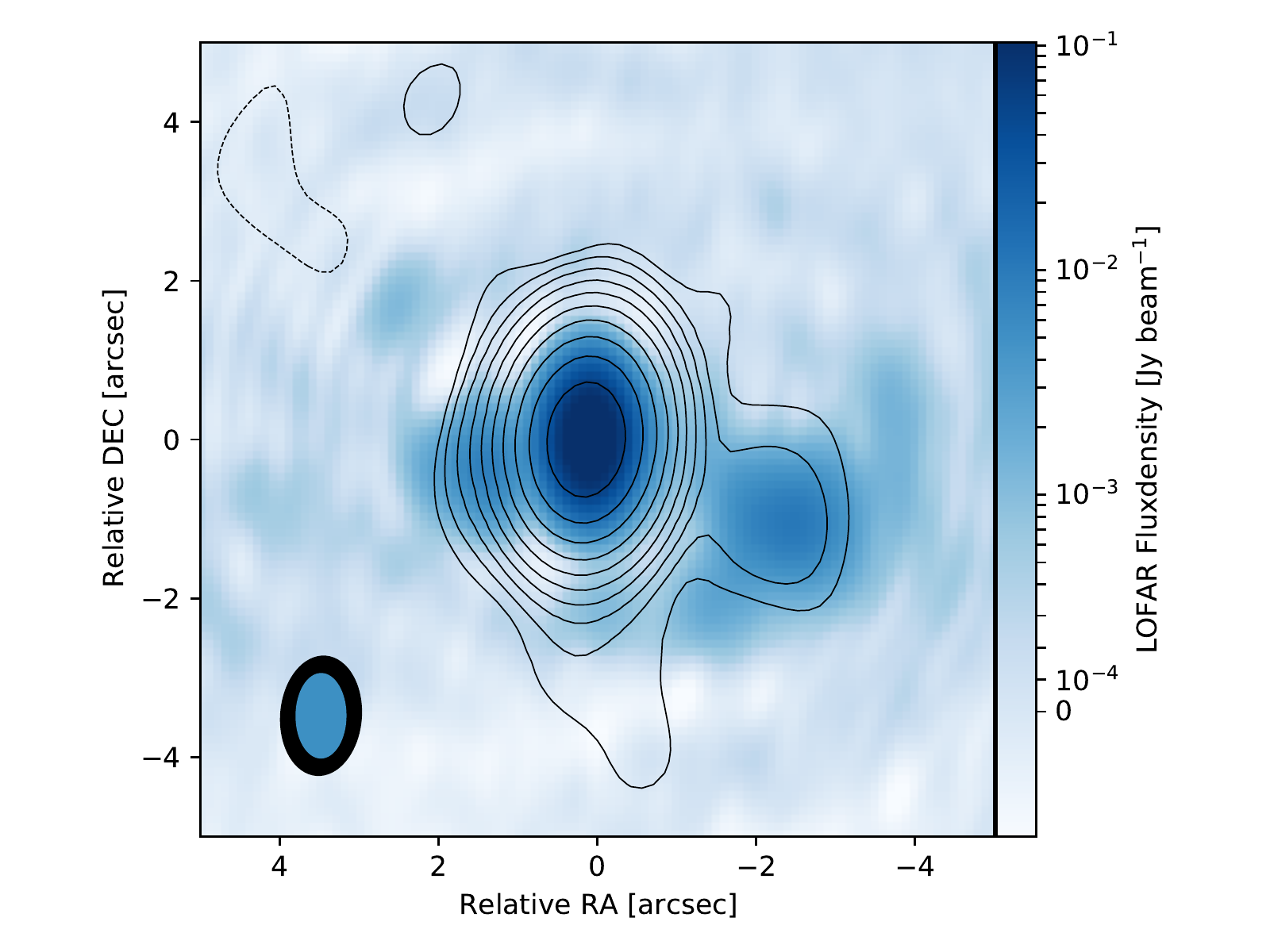}
        \caption{Colour-coded LOFAR image of GB\,1508+5714 at 144\,MHz with the 1.43\,GHz VLA contours from \cite{cheung2004} overlaid. The LOFAR and VLA restoring beams are shown in the bottom left corner as blue and black ellipses, respectively. The VLA contour levels are drawn at ($-1, 1, 2, 4, 8,$ ... ) times $0.23$\,mJy$\,$beam$^{-1}$. The RMS in th VLA data is $0.20$\,mJy$\,$beam$^{-1}$. The restoring beam size is $1.52'' \times 1.03''$ with a position angle of $-4.44^\circ$. The restored LOFAR beam size is $1.08'' \times 0.643''$ with a position angle of $-0.27^\circ$.
        }
        \label{fig:lofarvla}
\end{figure}

The final image\footnote{Fig. \ref{fig:lofarobs}  {\bf is available in electronic form at the CDS via anonymous ftp to cdsarc.u-strasbg.fr (130.79.128.5) or via \href{http://cdsweb.u-strasbg.fr/cgi-bin/qcat?J/A+A/}{http://cdsweb.u-strasbg.fr/cgi-bin/qcat?J/A+A/}}} is shown in Fig. \ref{fig:lofarobs}. 
Most of the flux density $(\sim 86\%)$  is contributed by the core component in the centre of the image. Significant bright and resolved emission is located east and west of the core, with the  western component appearing brighter than the eastern one. The largest extent of the whole structure is about 6" ($\sim 41$\,kpc).
Previous observations with the Karl G. Jansky Very Large Array (VLA) \citep{cheung2004} are in good agreement with our LOFAR image. For a direct comparison, Fig.~\ref{fig:lofarvla} shows the contour lines from the VLA observation at 1.43\,GHz overlaid on  the LOFAR brightness distribution.
The calibrated archival VLA data\footnote{\href{https://hea-www.harvard.edu/XJET/img-data.cgi?1508_vla_2ghz_1995jul.txt}{\text{\tiny{https://hea-www.harvard.edu/XJET/img-data.cgi?1508\_vla\_2ghz\_1995jul.txt}}}} came from a single five-minute snapshot observation 
in the A-configuration. This yielded a 10$\sigma$ detection of the western component containing about 1.2\,mJy. The core flux density amounts to $(224\pm11)$\,mJy at 1.43\,GHz at the time of the VLA observation.
Despite the longer wavelength, the ILT image features a better angular resolution than the VLA.
The eastern component shows a steep spectrum and is thus much fainter and appears unresolved at 1.43\,GHz.
The eastern component seen by LOFAR is not clearly detected by the VLA observation (see \citealt{cheung2004}). 
This component is not seen in the Chandra X-ray image either, even though the   resolution is comparable to that of  our ILT image \citep{mckeough2016}.

\begin{figure}[ht]
        \centering
        \includegraphics[width=\linewidth]{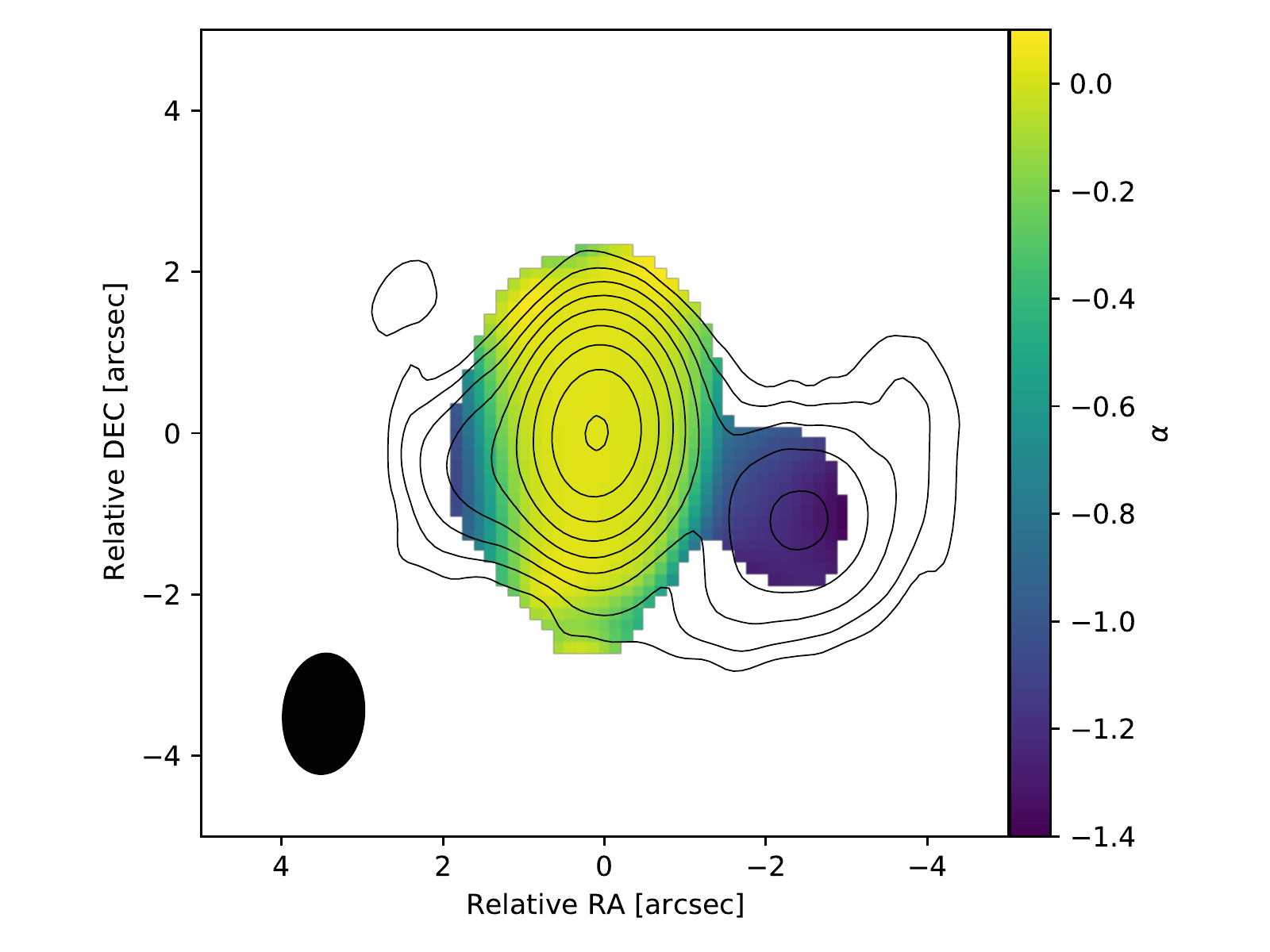}
        \caption{Colour-coded spectral-index image between 144\,MHz and 1.43\,GHz. The contour levels show the flux-density distribution at 144\,MHz with ($-1, 1, 2, 4, 8,$ ... ) times $0.78$\,mJy$\,$beam$^{-1}$.The joint beam (bottom left corner) used to restore  the LOFAR and the VLA images is $1.52'' \times 1.03''$ with a position angle of $-4.44^\circ$.
        }
        \label{fig:spixlofarvla}
\end{figure}

Reconvolving the LOFAR observation ({beam size $1.08'' \times 0.643''$ at position angle, p.a., of $-0.27^\circ$}) with the beam parameters of the VLA observation ({beam size $1.52'' \times 1.03''$ p.a. $-4.44^\circ$} see contours in Fig. \ref{fig:spixlofarvla}), we can see a very good visual agreement. The reconvolved LOFAR observation still shows a prominent resolved emission component to the east of the bright core. In comparison, the VLA observation only shows marginal evidence of an extension in this direction. 
This is consistent with the component featuring a steep spectral index.\newline
Combining the reconvolved LOFAR data with the VLA data, performing an alignment correction, and fitting a power law for each pixel with a flux value above a 2$\sigma$ threshold ($\sigma$ being $0.13$\,mJy$\,$beam$^{-1}$ for the ILT image and $0.20$\,mJy$\,$beam$^{-1}$ for the VLA image), we calculated the spectral index map shown in Fig.~\ref{fig:spixlofarvla}. Pixels for which the $2\sigma$ threshold is not met in either image are depicted in white. The core region shows a flat spectrum, which suggests  that the variable flat-spectrum core was in a similar emission state during the LOFAR and VLA observations.
We measure a steep spectral index\footnote{We define the spectral index $\alpha$ via $S_\nu \propto \nu^\alpha$, with $S_\nu$ radiative flux density.}  for the western component with a median value of -1.2  ($\text{min}=-1.4$, $\text{max}=-0.84$). \cite{cheung2006} reported a spectral index of \mbox{$(-1.4 \pm 0.2)$} between 1.4\,GHz and 5\,GHz, which does not indicate any significant spectral break between 144\,MHz and 5\,GHz. 
While the measured flux density of the eastern component in the VLA data is not significant enough to satisfy our flux threshold requirements, we can see a tendency of the same steepness, reaching $\alpha\lesssim-1.1$ at the easternmost edge of the spectral-index image. This value is an upper limit due to the non-detection by the VLA. Comparable values were reported previously for another high-redshift blazar in which subarcsecond-scale emission features close to the core were identified as hot spots \citep{kappes2019}. Considering that powerful quasars resemble rotated \cite{fanaroff1974} class II (hereafter FR-II) radio galaxies, in the unification model \citep{urry1995} observations of nearby FR-II radio galaxies at higher frequencies (accounting for  \textit{K}-correction) can be compared to our measurements. Here flatter spectral indices could be observed for hot spot regions \citep{ischwara-chandra2000,harwood2015}. 

Our LOFAR observation shows a total flux density of \mbox{$F_{\nu,\text{T}}=(254 \pm 33)$\,mJy.} To determine the parameters of both components (eastern and western), we use a one-Gaussian model (major axis equals minor axis) component in each corresponding region to model the emission (see crossed circles in Fig. \ref{fig:lofarobs}). We find \mbox{$F_{\nu,\text{W}}=(20.5 \pm 2.6)$\,mJy} for the western component  with a diameter of about 0.87'' ($\sim6\,$kpc), and \mbox{$F_{\nu,\text{E}}=(12.7 \pm 1.6)$\,mJy} for the eastern component  with a diameter of about 0.62'' ($\sim4\,$kpc). Similarly, we derive the flux density for the unresolved core component as \mbox{$F_{\nu,\text{core}}=(218 \pm 28)$\,mJy}. The residual flux density of \mbox{$F_{\nu,\text{R}}=(2.8 \pm 0.3)$\,mJy} remains unattributed in model components. The uncertainty on the residual flux density was calculated by considering the RMS in the residual image about 20''   from the source, in a box with an area of about 200 beams.
This RMS was normalised by the square root of the number of pixels, and subsequently converted from mJy$\,$beam$^{-1}$ to mJy. The total extended flux density is the sum of the three non-core emission contributions: \mbox{$F_{\nu,\text{ext}}=(36.0 \pm 3.1)$\,mJy.}

\section{Modelling}
\label{sec:modeling}
In this section we present the physical modelling done with our data as reference. This is necessary to determine the observational properties expected from a source with the same physical parameters but observed in the local Universe, where the CMB energy density is much smaller. These observational properties are plotted in our subsequent discussion, and indicated in  figures \ref{fig:blendedhotspot}, \ref{fig:lobe}, \ref{fig:hotspot}, and \ref{fig:weaklobe} as `no CMB'.

We focus on the flux density of the extended emission regions. In \cite{ghisellini2015} these regions were assumed  ad hoc  to emit between 1\% and 10\% of the  total jet power. We measure the value $\log_{10}(\nu F_{\nu\text{,ext}})=-16.30^{+0.06}_{-0.05}$ in cgs units. If the spectral index $\alpha$ and the flux density of a certain region are known, then the corresponding specific luminosity can be determined as follows \citep{condon1988}:
\begin{align}
\label{eq:lumin}
    L_\text{region} = \dfrac{F_\text{region} \cdot 4 \pi \cdot (c\cdot z)^2}{H_0^2 \cdot (1+z)^{1-\alpha_\text{region}}}. 
\end{align}

Assuming $\alpha=0$ for the core, we derive $\log_{\rm 10} (L_{\text{144\,MHz}, \text{core}})=27.2$ (in cgs units).
Spectral index values can vary across extended regions, so we use the minimum (-1.4) and maximum (-0.84) values as boundaries to calculate a range of intrinsic luminosities, accounting for the $K$-correction using the spectral index. The luminosity range for the extended emission is $\log_{10}(L_{\text{144\,MHz},\text{ext}})=25.4$ to $25.8$ (in cgs units),  
corresponding to 1.6\% to 4.0\% of the core luminosity.
Following the canonical classification in \cite{fanaroff1974}, the luminosity we calculate for GB\,1508+5714 corresponds to the case of a FR-II radio galaxy \citep{owen1994}. The measurements of \citet{owen1994} are taken at 1.4\,GHz for nearby sources (relative to \object{GB\,1508+5714}) and so we must take the \textit{K}-correction into account to compare results. Our observed frequency of about 144\,MHz corresponds to an emitted frequency of about 760\,MHz, well below 1.4\,GHz. Considering that the transition luminosity between FR-I and FR-II increases with decreasing frequency \citep{kembhavi1999} we remain above the FR break: the FR-II association for \object{GB\,1508+5714} remains valid.
\object{GB\,1508+5714} belongs to the class of flat spectrum radio quasars (FSRQs) which are considered the beamed counterparts of FR-II radio galaxies \citep{orr1982} within the AGN unification scheme \citep{peacock1987,scheuer1987,barthel1989,urry1995}.
With recent findings \citep{mingo2019} on this canonical classification showing that FR-II radio galaxies can be up to three orders of magnitude below the traditional FR break, FR-I radio galaxies seem to be more strictly defined by lower luminosities; this would also suggest that GB\,1508+5714 would more appropriately correspond to an FR-II object than an FR-I.

The broad-band spectral energy distribution (SED) of FSRQs is 
characterised by thermal emission (i.e. an accretion disc in the optical--UV, a torus in the IR, 
and possibly a hot corona in the X-ray band)  and non-thermal emission produced by a jet pointing
at a small angle to the line of sight, and thus strongly boosted by relativistic beaming.
FSRQs are typically strongly variable at different wavebands, and simultaneous variations are not uncommon \citep{meyer2019,shukla2020,acciari2020,kramarenko2021,acciari2021}. The emission is therefore typically modelled as coming from a single region.
In single-zone models, the emitting region must not be located too close to the accretion disc to avoid being too compact and thus absorbing all the $\gamma$-ray emission in the
$\gamma$--$\gamma \to$ e$^+$ e$^-$ process. Similarly, it cannot be too far from the accretion disc, to make it possible to explain the observed fast variability. 
This gives boundaries for the location of the source of emission at approximately $\sim10^3 - 10^4$   Schwarzschild radii from the black hole \citep{liu2006}.
All blazar SEDs exhibit two humps: one at low energy, due to synchrotron emission,
and another at high energy, due to inverse Compton processes (although some
hadronic processes may also contribute). 
In very powerful sources, such as GB\,1508+5714, the first hump peaks in the sub-millimetre band, and the high-energy sources peak in the $\sim$MeV band.
The last observable quantity is generally the accretion disc emission, peaking in the UV.
In these sources we can thus model the disc emission to find estimates for its luminosity (and hence the accretion rate)  and its black hole mass, independently of other methods such as emission line widths and luminosities.
The uppermost data points in Fig.~\ref{fig:blendedhotspot} show the SED  produced by the accretion disc, the molecular torus and the emission from that part of the jet thought to produce most of the non-thermal radiation we see. Unfortunately, we do not have measurements in the appropriate IR band to constrain the molecular torus emission properties. {\bf The model, indicated as `no CMB', shows the expected SED in the absence of the CMB. It can be seen that in this case the radio flux density would be enhanced and the X-ray emission severely depressed. This shows how a source with the same parameters would appear if it were nearby, where the CMB energy density is much lower (i.e. by a factor $(1+z)^4 \sim 790$).}

It should be noted that the core-dominated radio spectrum is flat (i.e. $F_\nu \propto \nu^0$, dashed blue line in Fig. \ref{fig:blendedhotspot}) down to 50 MHz \citep{deGasperin2021}.
This emission is thought to be the superposition of the flux  produced by larger and larger 
jet emitting zones, characterised by a self-absorption frequency $\nu_{\rm t}$
inversely proportional to  the emitting size \citep{blandford1979}.
In the case of a conical jet, we should approximately have $\nu_{\rm t} \propto 1/R_{\rm j}$,
where $R_{\rm j}$ is the distance from the black hole.
Since the Doppler boost varies strongly with the viewing angle, we can conclude that the jet does not bend significantly, at least up to the regions producing the 50 MHz flux, otherwise the 50 MHz flux would not lie on the $\nu^0$ line.

This implies $R_{\rm j}(50\, {\rm MHz})\sim 1.3$ kpc if we assume that the flux at 300 GHz is produced in the jet region emitting the rest of the jet spectrum.
All the jet parameters, which are listed in Table \ref{parajet} for  convenience, are very similar to the blazars detected by the Large Area Telescope (LAT) on board the {\it Fermi} satellite \citep{atwood2009} and analysed in \cite{ghisellini2015b}. The jet power ($P_{\rm jet} \sim 3\times 10^{47}$ erg s$^{-1}$) is dominated by the bulk motion of cold protons (assumed to be equal in number to the emitting electrons, i.e. assuming no significant contribution from electron--positron pairs).
The lower limit to $P_{\rm jet}$ is $10^{46}$\,s\mbox{erg s$^{-1}$}, which is the total power in radiation emitted by the jet. 
\\
\\
We note that the `jet--hot spots--lobe' models are a simplified representation of a likely complex reality. In particular, the lobe and the hot spots are idealised to be spheres homogeneously filled with tangled magnetic fields and relativistic particles, without internal gradients. When dealing with extended sources at high redshifts, the contribution of the CMB becomes important since it can be the dominant source of seed photons to be scattered at high energies. Relativistic electrons can inverse Compton scatter their own synchrotron photons, called synchrotron self-Compton (SSC), and also the CMB photons. 
In the latter case we call this the  external Compton process. The energy density of the CMB increases with redshift as $(1+z)^4$, and we have equipartition between the magnetic ($U_B$) and the CMB ($U_{\rm CMB})$ energy densities for a magnetic field value $B_{\rm eq}$ given by
\begin{equation}
B_{\rm eq}\, =\, 3.26\times 10^{-6} (1+z)^2 {\rm G} \, =\,  
9.2 \times 10^{-5} \left[\frac{(1+z)}{1+4.309}\right]^2   
\,\,\,\, {\rm G}.
\end{equation}
For magnetic fields smaller than $B_{\rm eq}$, we have $U_{\rm CMB} > U_B$, 
and the high-energy (typically X-ray) luminosity
produced by the external Compton is greater than the synchrotron luminosity.\\
A diffuse extended component 
has been measured in X-rays by Chandra \citep[][ butterfly spectrum in the figure \ref{fig:blendedhotspot}]{siemiginowska2003, yuan2003}.
Assuming that this X-ray component is produced by the external Compton process, 
the ratio of this X-ray luminosity to the extended radio luminosity is 
a measure of the ratio $U_{\rm CMB}/U_B$. 
As illustrated in Fig. \ref{fig:lobe}, the ratio is between two and three orders of
magnitude.
We therefore can find an upper limit for the value of the magnetic field of the extended (hot spots and/or lobe{\bf s}) components $B_{\rm ext}$
\begin{equation}
\frac{L_{\rm X, ext}}{L_{\rm syn}} \sim  \frac{U_{\rm CMB}}{U_B}  \to 
B_{\rm ext}\, \lesssim   \, \left[\frac{8\pi U_{\rm CMB}}{L_{\rm X, ext}/L_{\rm syn}}\right]^{1/2}  
\label{b}
,\end{equation}
giving $B_{\rm ext} \lesssim 3$ $\mu$G for $L_{\rm X, ext}/L_{\rm syn}=10^3$.
This estimate is independent of the volume of the region emitting the diffuse X-rays.
Therefore, it can be applied to both the lobe and the hot spots,
depending on which structure is responsible for the diffuse X-ray flux.

The power in relativistic electrons $P_{\rm e}$ injected into the hot spots 
and the lobes  can be derived by requiring that they produce the observed X-ray luminosity $L_{\rm X, ext}$. As long as the radiative cooling is dominated by the CMB, $P_{\rm e}$ is also independent from the volume of the lobe. Some subtleties arise from the fact that we do not assign a given and fixed particle distribution, but rather find it self-consistently through a continuity equation that accounts for the cooling terms (synchrotron, SSC, external Compton, adiabatic).
Therefore, we specify a priori the total power injected in the relativistic electron population and its shape (a broken power law, very hard at low energies and steeper after a break), and hence find the emitting distribution. 
This procedure is crucial if we want to establish the effects of the CMB. When the CMB dominates the radiative cooling, it steepens the particle
distribution above the energy cooling break $E_{\rm cool}$, corresponding to the synchrotron frequency $\nu_{\rm cool}$. Above this frequency the synchrotron emission is depressed. We refer the reader to \cite{Ghisellini2014} and  \cite{ghisellini2015} for a complete description and the relevant equations of the model.

\section{Discussion}
\label{sec:discussion}
In this section we compare our observations to the predictions reported in the literature \citep{Ghisellini2009,Ghisellini2014,ghisellini2015} regarding the proposed CMB quenching mechanism that was suggested to explain the apparent sparseness of high-redshift $(z \gtrsim 4)$ misaligned jetted AGNs in radio surveys. For completeness, a further three models are presented in the Appendix; these were excluded in the process of our analysis in favour of the one presented here. All these physical scenarios for \object{GB\,1508+5714} involve a FR-II-like source structure (pointed at a small angle to the line of sight) with varying lobe and hot spot contributions to the observed radio and X-ray data. While it is straightforward to identify the core of the source, the two components observed east and west of the core are considered extended lobes. The brightest points in these lobes likely contain the hot spots,
but their physical sizes might well be smaller than the beam size in our observation\footnote{For a simple conical jet model, we would expect a hot spot diameter of about 2.5\,arcsec, assuming a jet inclination angle of 3$^\circ$ and an opening angle of 2.5$^\circ$. For larger inclinations and/or smaller intrinsic opening angles (e.g. \cite{pushkarev2009jet}) this can yield substantially smaller angular scales.}, and it is unclear what fraction of the emission should be attributed to lobe emission versus hot spots. 
There are cases where they might be interpreted as two individual hot spots, such as \citet{kappes2019}, but the situation was clearer in that case than for \object{GB\,1508+5714}.
The additional residual radio flux density measured by LOFAR (amounting to $F_{\nu,\rm R}=$2.8 mJy) is estimated by taking the difference between the total flux density and the flux densities of the resolved components. This can be attributed to artefacts introduced during calibration, or non-Gaussian extensions of the lobes or hot spots. Artefacts are commonplace in radio-interferometric images, and so we proceed with the postulate that these extensions are not real features. For the sake of completeness, the possibility that they are is explored in other possible models (denoted B, C, and D) in Appendix \ref{app:discussion}. We therefore refer to the model in which the non-Gaussian extensions are dismissed as not real features as model A.
$~$
\newline\newline

\begin{figure}[ht]
        \centering
        \includegraphics[width=\linewidth]{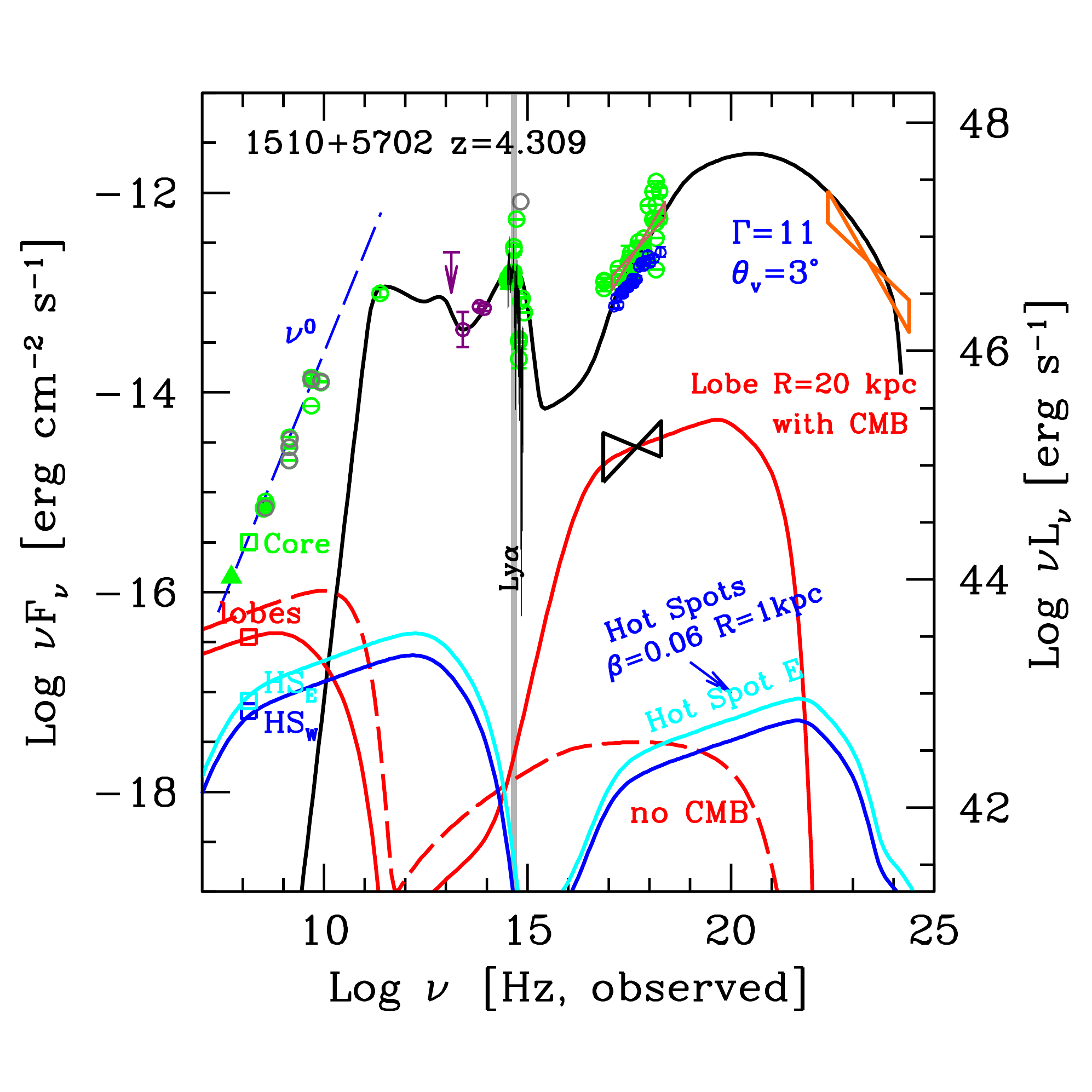}
       \caption{SED of 1510+5702, from radio to $\gamma$-rays. 
       The model shown (see  parameters in Table \ref{parajet}) 
       describes the non-thermal jet emission, the accretion disc, and the molecular torus contributions (top black solid). All data are archival (https://tools.ssdc.asi.it/). For the hot spots and the lobes  (as labelled; see corresponding model parameters in Table \ref{paralobe}) this figure corresponds to model\,A. The lobe with a radius of 20\,kpc is responsible for the diffuse X-ray emission observed by Chandra (black butterfly spectrum). The dashed red line is the lobe emission if there were no CMB. The vertical grey line gives the position of the Ly$\alpha$ line. The dashed blue line is not a fit of the radio spectrum of the core, but rather  has been drawn to guide the eye. The labelled data points at 144~MHz are the LOFAR determinations from this work.
       }
        \label{fig:blendedhotspot}
\end{figure}

{\bf Model\,A ---} We consider that the two lobes dominate the western and eastern components and contain a faint unresolved hot spot each. We limit the maximum possible brightness of the two hot spots to  $\lesssim 4.3$\,mJy and $\lesssim 5.3$\,mJy, respectively, in order to not exceed the local surface brightness of their two lobes. The total flux density for both lobes accounts for 23.6\,mJy. In this model the two strong lobes dominate the X-ray emission (although the Chandra image suggests a more efficient X-ray production associated with the western lobe). Figure~\ref{fig:blendedhotspot} shows the SED for this model. In this case the two strong lobes are dominating both the total extended radio flux east and west of the core and the diffuse X-ray flux. As listed in Table~\ref{paralobe}, the lobes are not very far from equipartition ($E_{\rm e}/E_{\rm B}\sim 89$) and the power required to energise this structure is $5\times 10^{46}$ erg s$^{-1}$. The energetics of the hot spots are consistent with a state of equipartition.

$~$
\newline\newline
%

$~$
Although we favour model\,A, which is a self-consistent and physically motivated representation of the observed properties of GB\,1508+5714, it does not result in the tightest constraints on hot spot luminosities nor does it allow us to derive constraints on hot spot advance speeds, which depends on differences in Doppler boosting. This was a motivation to considering the models B, C, and D shown in the Appendices; the relevant calculations that derive from the best-fit values for these models are given in Appendix\,\ref{app:speed} and Appendix\,\ref{app:age}. Properly constraining these models, however, will require creating  higher-quality interferometric images than was possible  with the quality of the data available to us; they are therefore only given as a point of reference, and we do not consider them to have scientific value at this time. In all cases, we can state that CMB quenching must be present as expected to explain the observations.

\begin{table*} 
\centering
\begin{tabular}{l l l l l l l l l l l l l l l l l l}
\hline
\hline
Comp &$z$ &$M$ &$L_{\rm d}$ &$R_{\rm diss}$ &$R_{\rm BLR}$ &$P^\prime_{\rm e, jet, 45}$  &$B$ &$\Gamma$ &$\theta_{\rm V}$  
  &$P_{\rm jet, 45}$ \\ 
~[1] &[2] &[3] &[4] &[5] &[6] &[7] &[8] &[9] &[10] &[11]   \\
\hline   
jet &4.309 &1.5e9  &32  &630 &560 &0.03 &1.9  &11 &3 
&360  \\   
\hline
\hline 
\end{tabular}
\vskip 0.4 true cm
\caption{
Adopted parameters for the jet model shown in Fig. \ref{fig:blendedhotspot}--Fig. \ref{fig:weaklobe}.
Col. [1]: Component; 
Col. [2]: redshift;
Col. [3]: black hole mass in units of solar masses;
Col. [4]: disc luminosity in units of $10^{45}$ erg s$^{-1}$;
Col. [5]: distance of the dissipation region from the black hole in units of $10^{15}$ cm;
Col. [6]: size of the broad line region in units of $10^{15}$ cm;
Col. [7]: power injected in the jet in relativistic electrons, calculated in the comoving 
frame, in units of $10^{45}$ erg s$^{-1}$;
Col. [8]: magnetic field in G;  
Col. [9]: bulk Lorentz factor;
Col. [10]: viewing angle in degrees;
Col. [11]: total kinetic plus magnetic jet power in units of $10^{45}$ erg s$^{-1}$.
The values of the powers and the energetics refer to one jet.
We assume,  injected throughout the source, $Q(\gamma)$  relativistic electrons with broken power law distribution,
i.e. $Q(\gamma) \propto \gamma^{-1.5}$ below $\gamma=200$ and $Q(\gamma)\propto \gamma^{-3}$ between $\gamma=200$
and $\gamma=4000$. 
}
\label{parajet}

\end{table*}

\begin{table*} 
\centering
\begin{tabular}{l l l l l l l l l l l l l l  l l l}
\hline
\hline
Model &Comp. &$R$ &$\theta_{\rm v}$ &$\beta$ &$P_{\rm e,45}$ &$B$  &$\gamma_{\rm b}$ &$\gamma_{\rm max}$ 
&$\log E_{\rm e}$  &$\log E_{\rm B}$  &$E_{\rm e}/E_{\rm B}$ \\ 
~[1] &[2] &[3] &[4] &[5] &[6] &[7] &[8]&[9] &[10] &[11] &[12]\\  
\hline   
A &HS   &1   &$3^\circ$, $-177^\circ$ &0.06 &2.3  &180  &600  &1e6   &56.2 &56.2 &0.9   \\
A &lobe &20  &--                      &0    &50   &7    &300  &1e5   &59.2 &57.3 &89   \\
B &HS   &2   &$3^\circ$, $-177^\circ$ &0.06 &5.4  &140  &600  &1e6   &56.9 &56.9 &1  \\  
B &lobe &20  &--                      &0    &50   &2.4  &300  &1.e5  &59.2 &56.4 &630  \\
C &HS   &2.9 &$3^\circ$, $-177^\circ$ &0.06 &600  &7    &400  &1.e4  &59.4 &54.8 &4e4\\
D &HS   &2.9 &$3^\circ$, $-177^\circ$ &0.06 &600  &7   &400  &1.e4 &59.4  &54.8 &4e4 \\
D &lobe &20  &--                      &0    &1    &12   &400  &1e4   &57.8 &57.7 &1.2  \\
\hline
\hline 
\end{tabular}
\vskip 0.4 true cm
\caption{
Adopted parameters for the hot spot and lobe models shown in Fig. 
\ref{fig:blendedhotspot}--Fig. \ref{fig:weaklobe}.
Col. [1]: Type of model 
Col. [2]: component (HS = hot spot); 
Col. [3]: size in kpc;
Col. [4]: viewing angle;
Col. [5]: bulk velocity;
Col. [6]: Power injected in relativistic electrons in units of $10^{45}$ erg s$^{-1}$;
Col. [7]: magnetic field in $\mu$G;
Col. [8] and Col. [9]: break and maximum Lorenz factor of the injected electron distribution;
Col. [10]: logarithm of the total energy in relativistic electrons, in erg;
Col. [11]: logarithm of the total energy in magnetic field, in erg.
Col. [12]: Total energy in relativistic electrons over  total energy in magnetic field.
The values of the powers and the energetics refer to  {\it each} hot spot and lobe.
The lobe flux shown in the figures corresponds to {\it two} lobes.
We assume to inject throughout the source $Q(\gamma)$  relativistic electrons with a broken power law distribution,
i.e. $Q(\gamma) \propto \gamma$ below $\gamma=\gamma_{\rm b}$ and $Q(\gamma)\propto \gamma^{-2.7}$ between 
$\gamma_{\rm b}$ and $\gamma_{\rm max}$. 
}
\label{paralobe}
\end{table*}

\section{Conclusion}
\label{sec:conclusion}
New high-resolution images taken by the ILT allow  a novel view on the high-redshift blazar GB\,1508+5714, showing a previously unseen component in the eastern direction. By reconvolving the ILT data with the VLA beam, we were able to create a spectral index map, from which we derive constraints on the spectral indices of individual components: $-1.2^{+0.4}_{-0.2}$ for the western component; steeper than $-1.1$ for the eastern region; $0.023\pm0.007$ for the core.
We then considered a model where the hot spots are unresolved in the detected components and are blended by the lobe emission. The X-ray emission originates from the lobes and is partially (eastern lobe) blended by the bright core region. For completeness we  considered three other possible interpretations that are worked out in Appendix \ref{app:discussion}. The preferred model's results are consistent with hot spots in a state of equipartition, and lobes nearly so. 
In all models we can confirm the necessity of CMB quenching processes as proposed by \cite{ghisellini2015}. 
We note the need for more observations of these sources, especially with the current more mature ILT configuration, which features more international stations, resulting  in a more sensitive instrument, but more importantly in a much more stable instrument overall. This will allow future observations to result in much higher-quality data than was possible for the ILT to create in 2015 when our observation was taken; in turn, it could make a full investigation and characterisation of the models discussed in the Appendix possible as they were   dismissed primarily due to problems with data quality.
Nevertheless, in this paper we have shown that it is possible to study high-z blazars with the 2015 configuration of the ILT; any observations performed after the time of publication can only provide superior observational constraints. In the future, expanding the sample of high-z blazars resolved in multi-frequency will allow us to conduct a statistical study of the population.


\begin{acknowledgements}
    We thank the anonymous referee for the constructive comments, which helped us to improve the manuscript. We are grateful to  Laura Vega-Garc\'ia for insightful discussions and useful comments on the manuscript.
    P.R.B. wants to acknowledge support from the 'DATEV-Stiftung Zukunft'  through funding the interdisciplinary data lab 'DataSphere@JMUW'.
    Work by C.C.C. at the Naval Research Laboratory is supported by NASA DPR S-15633-Y. MP acknowledges support by the Spanish Ministry of Science through Grants PID2019-105510GB-C31, PID2019-107427GB-C33, and from the Valencian Government through grant PROMETEU/2019/071.
    A.D. acknowledges support by the BMBF Verbundforschung under the grant 05A20STA.
    J.M. acknowledges financial support from the State Agency for Research of the Spanish MCIU through the ``Center of Excellence Severo Ochoa'' award to the Instituto de Astrof\'isica de Andaluc\'ia (SEV-2017-0709) and from the grant RTI2018-096228-B-C31 (MICIU/FEDER, EU).
    E. B. acknowledges support from the ERC-Stg grant DRANOEL, n.714245.
        
\end{acknowledgements}

\bibliographystyle{aa}


\begin{appendix}
\section{Discussion for models B, C, and D}
\label{app:discussion}

\begin{figure}[ht]
        \centering
        \includegraphics[width=\linewidth]{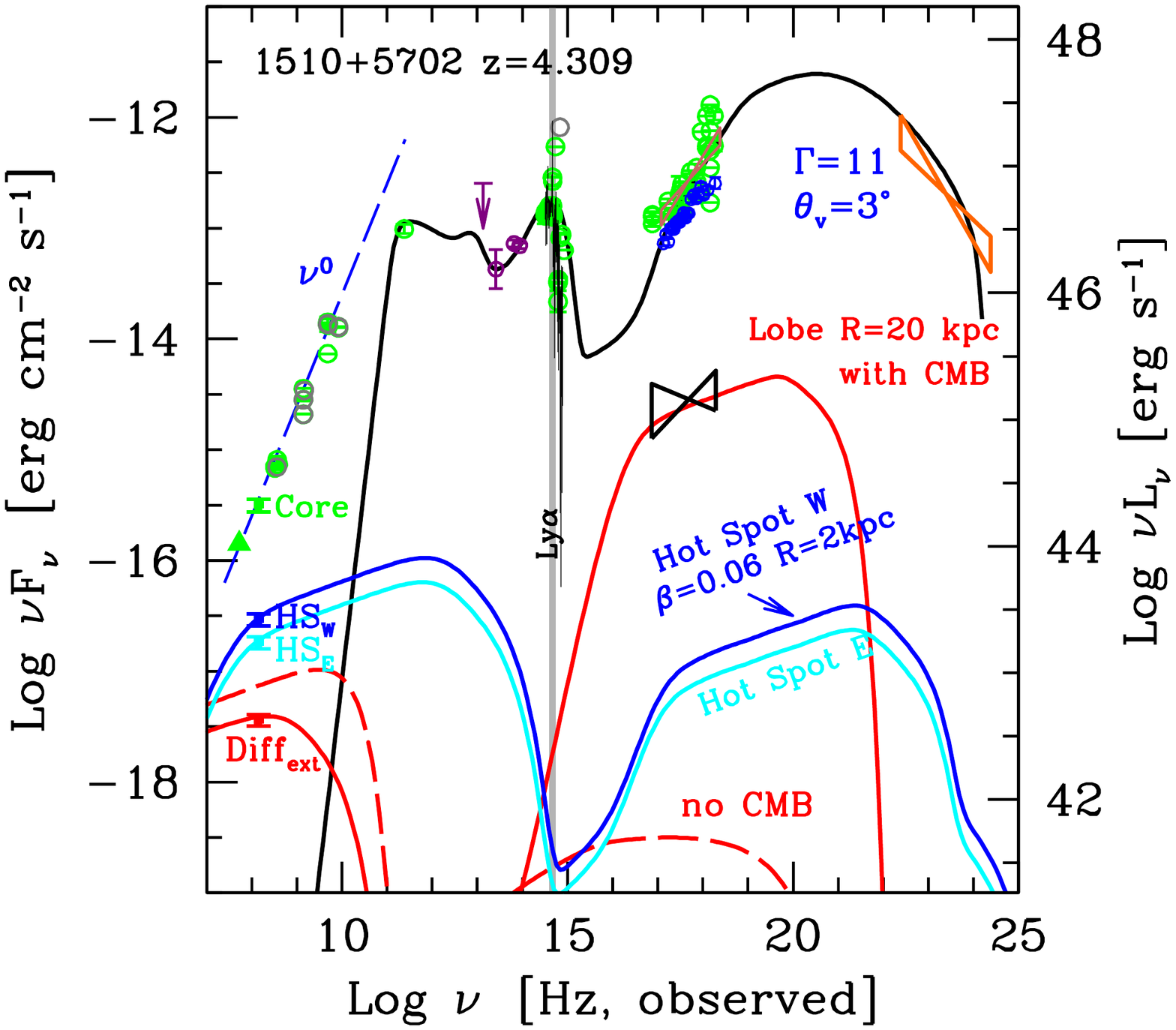}
       \caption{
       As Fig.~\ref{fig:blendedhotspot}, but for model\,B. The hot spots are 2\,kpc in size with a weaker magnetic field. All parameters are given in Table~\ref{paralobe}, together with those adopted for the lobes.
      }
        \label{fig:lobe}
\end{figure}

\begin{figure}[ht]
        \centering
        \includegraphics[width=\linewidth]{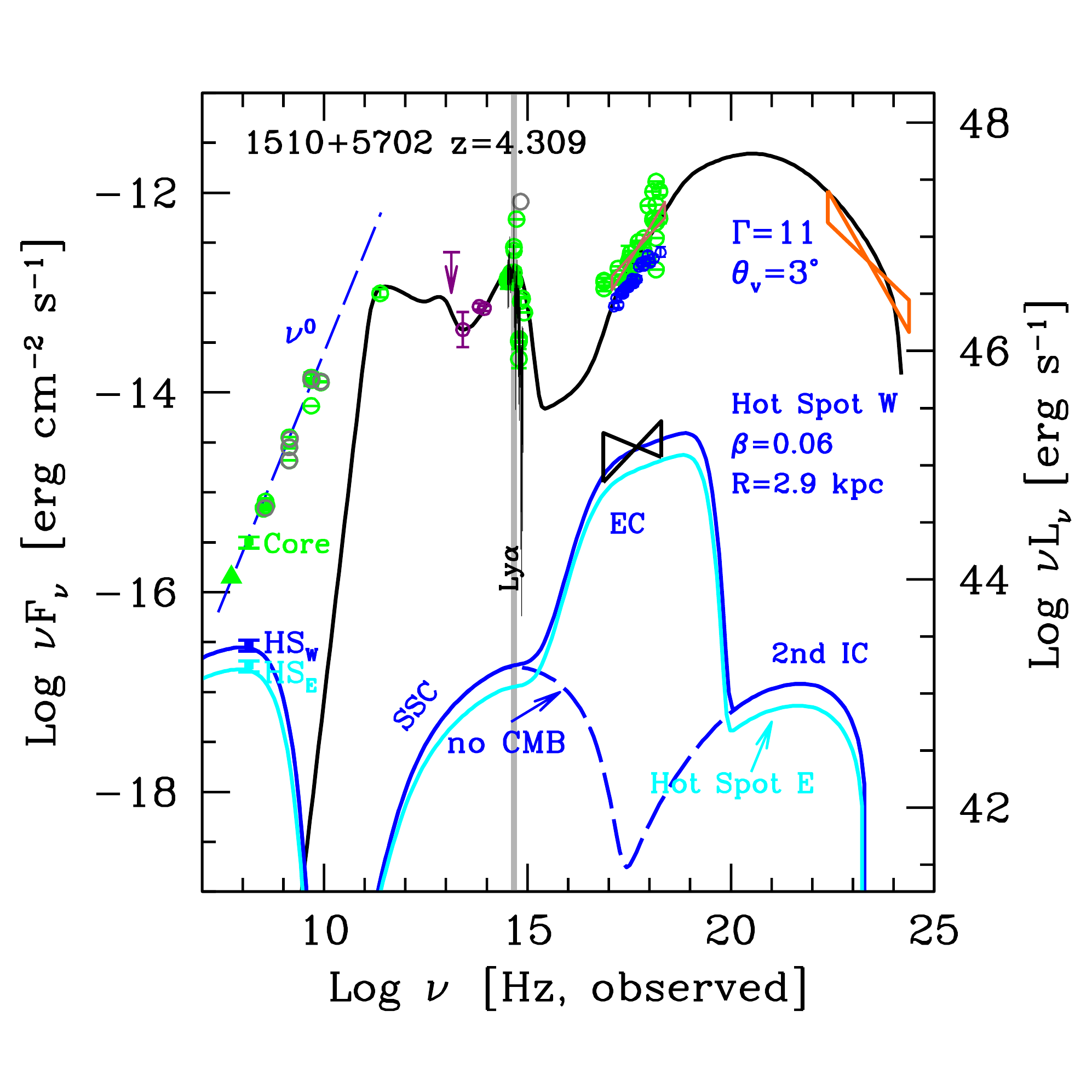}
       \caption{As Fig.~\ref{fig:blendedhotspot}, but for model\,C. The diffuse X-ray emission observed by Chandra is produced by the hot spots, and there are no lobes.
       We also show how the flux of the western hot spot would appear if 
       there were no CMB. This shows that most of the X-ray diffuse flux is the result of external Compton process with the CMB photons, while the SSC component is very weak. At high energies 
       the flux is dominated by the second-order Compton scatterings 
       (second IC).
       The parameters adopted for the hot spots are given in Table~\ref{paralobe}.
       }
        \label{fig:hotspot}
\end{figure}

\begin{figure}[ht]
        \centering
        \includegraphics[width=\linewidth]{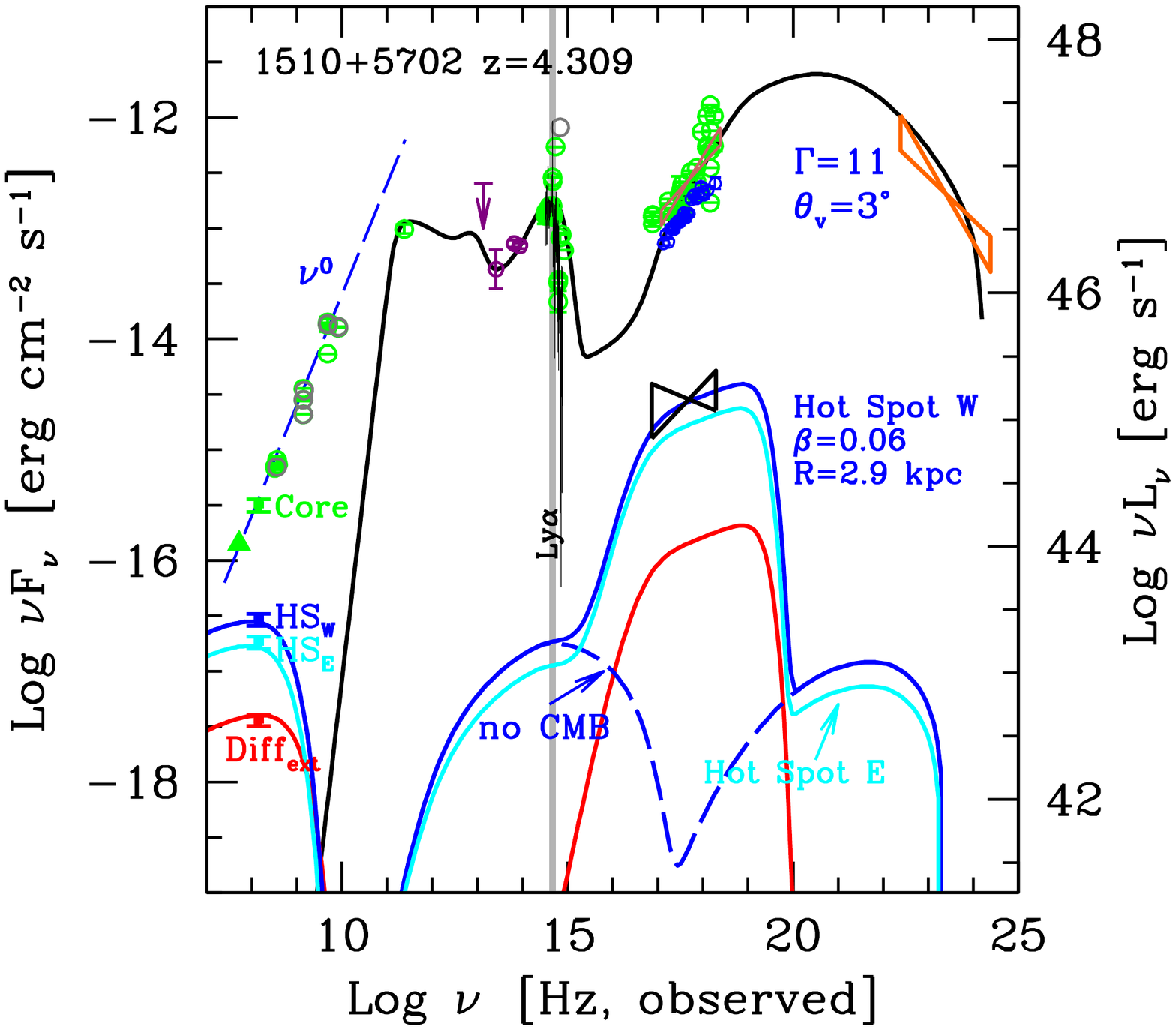}
       \caption{As Fig.~\ref{fig:blendedhotspot}, but for model\,D. The diffuse X-ray emission observed by Chandra is produced by the hot spots, but we assume that the lobes exist, and emit 2.8\,mJy flux density at 144\,MHz. We then study the parameters that a 20\,kpc lobe 
       must have in order to produce the radio, but not the observed X-rays.
       The hot spots have the same parameters as in Fig. \ref{fig:hotspot} and are given in Table~\ref{paralobe}, together with those adopted for the lobes.
       }
        \label{fig:weaklobe}
\end{figure}

Following the discussion in section \ref{sec:discussion} we present the three models that we were able to discard in the process of our scientific analysis of the data, but nevertheless find valuable to discuss for the sake of completeness. These models do not make  clear a priori assumptions on whether the hot spots or the lobes dominate the observed X-ray emission. With the parameters derived from the individual morphological interpretations and the modelling of the accretion disc, jet, hot spots, and lobes we present the results for the individual models B, C, and D.
$~$
\newline\newline
{\bf Model\,B ---} Here we assume that the residual 2.8\,mJy radio flux density is produced by two lobes, each with a radius of 20\,kpc (40\,kpc corresponds to the largest extent of the source in Fig.~\ref{fig:lofarobs}).
We further assume that these lobes produce the diffuse Chandra X-ray emission. We interpret the resolved components in the east and west as hot spots. Due to the assumption we have made, it follows that this model excludes the possibility that the hot spots contribute to the diffuse X-ray flux in any capacity. The corresponding SED and model are shown in Fig. \ref{fig:lobe}.
As discussed above (Eq. \ref{b}), the magnetic field strength must be less than the order of $\sim$microGauss, regardless of the size of the lobes. This very small value, together with the requirement to produce a large diffuse X-ray luminosity, results in best-fit models that can only be far from equipartition.
In our case $E_e/E_B \sim 630$.
The impact of the CMB\,  can be seen by examining the SED of the lobe in the absence of the CMB.
The first effect is a huge reduction of the X-ray flux, due to the absence
of seed CMB photons. The radio spectrum is also drastically different 
because the particle distribution would have a radiative cooling break at much higher energies. 
Therefore, the emission slope in the radio band would continue to be flat
up to the sub-millimetre band.  \\ \\
The two hot spots have similar, but not identical, fluxes.
{\bf Model\,B} allows us to explain this by invoking the different Doppler boosts due to the bulk velocity of the hot spots, assumed equal in all respects but the viewing angle.
We thus model them both as having a physical radius of 2\,kpc.
If the hot spots do not contribute significantly to the diffuse X-ray flux, we cannot find a unique solution. 
This strengthens the hypothesis made in model\,A that there should be equipartition between the relativistic electron and the magnetic field total energy in the hot spot. This constraint, listed in Table \ref{paralobe}, is also a strong enough constraint to fix the value of the power injected in each lobe, which amounts to $\sim$1/65 of the total jet power (emitted synchrotron radiation in our respective radio band).
For consistency with the viewing angle of the jet, we  assume 
$\theta=3^\circ$ and $\theta=180^\circ-3^\circ$ for the western and eastern
hot spot, respectively \citep{ghisellini2015}. 
\newline\newline
{\bf Model\,C ---} Here we assume that the lobes do not contribute at all to the observed radio emission (i.e. that they emit only below the ILT sensitivity threshold).
The diffuse X-ray flux is then produced entirely by the eastern and western components identified as hot spots, and the residual emission in the image is entirely unphysical.
We note that the Chandra image \citep{siemiginowska2003, yuan2003} shows a resolved component corresponding to the 
western hot spot, but no corresponding counterpart in the east.
This discrepancy could be explained if the eastern hot spot's X-ray emission were blended with the dominant core emission. The SED for this model is shown in Fig.~\ref{fig:hotspot}.
The hot spots' magnetic field value is then 7\,${\mu\text{G}}$,
which implies that the magnetic field is very far from equipartition: $E_e/E_B \sim 4\times 10^4$.
The power that the jet must supply to the hot spot in the form of 
relativistic electrons is very large, amounting 
to $P_{\rm e}\sim 6\times 10^{47}$ erg s$^{-1}$, even larger than $P_{\rm jet}$.
This is due to the fact that even the highest energy electrons cannot radiatively cool on a timescale shorter that the adiabatic timescale $\sim R/c$. The emission process is therefore very inefficient, requiring a large number of relativistic electrons (and thus a large injected power), leading
to a case very far from equipartition. 
\newline\newline
{\bf Model\,D ---} Here we consider that the lobes are responsible for the 2.8\,mJy residual flux density, but that the bulk of the extended radio emission is to be attributed to two strong hot spots.
In contrast to model B, the lobes do not contribute significantly to the diffuse X-ray emission, which is assumed to be dominated by the hot spots.  Figure~\ref{fig:weaklobe} shows the SED for this model.
This allows it to be close to equipartition; however, the problems associated with the hot spot properties of model\,B remain.
\newline\newline

In conclusion, model\,A is the only one that simultaneously satisfies the equipartition condition for the hot spots and the lobes near them. It is also by far the least energetically demanding model. Model\,B only reaches equipartition within the hot spots, but not the lobes; models\,C and D do not reach equipartition anywhere. This is the basis on which all three of the models described in this section are dismissed in favour of model\,A.

%

%

\section{Hot spot advance speeds}
\label{app:speed}
\begin{figure}[h]
        \centering
        \includegraphics[width=\linewidth]{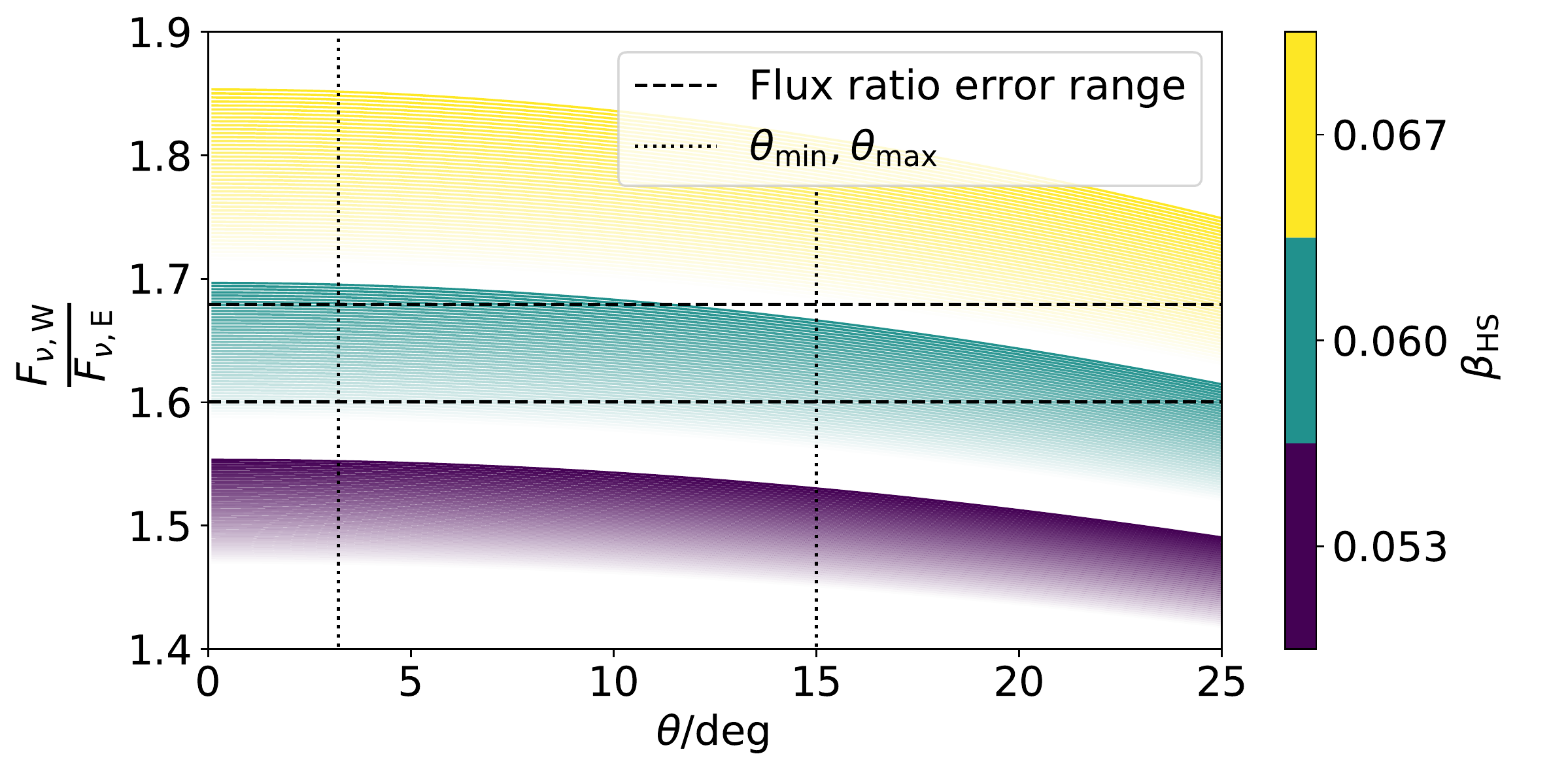}
        \caption{Ratio of the flux of western to eastern components, $R$, as a function of the inclination angle, $\theta$, and the hot spot speed, $\beta_{HS}$. The  different hot spot speeds are colour-coded;  the spectral index is given by the opaqueness of the different  regions   (a more saturated tone is a steeper  $-0.84 > \alpha > -1.4$). The dashed horizontal lines constrain the parameter space with $R$ within its error (see Eq. \ref{eq.:R}).The $\theta_{\mathrm{min}}$ value is determined by comparing the measured projected size with large radio galaxies. The $\theta_{\mathrm{SED}}$ value shows the suggested inclination angle of the inner jet by the SED.
        }
        \label{fig:parameter_space}
\end{figure}

Although models B, C, and  D must be dismissed because of  their failure to satisfy equipartition conditions, they have the very attractive ability (if applicable) to use their distinction between hot spot and lobe emission to estimate the advance speed of the hot spot components. In this section we describe what the current results of such an analysis would be in order to demonstrate what could become possible if future observations are able to provide data of sufficient quality to constrain these models without relying on equipartition conditions. Parameters such as the hot spot speed and the jet inclination angle can be constrained using the ratio of the hot spot flux densities if we use the simplified model of a symmetric two-sided jet:
\begin{align}
   R =  \frac{F_{\nu,\text{W}}}{F_{\nu,\text{E}}} = \left(\frac{1+\beta\, \mathrm{cos}(\theta)}{1-\beta\, \mathrm{cos}(\theta)}\right)^{3-\alpha} \quad.
\label{eq.:R}
\end{align}
\\
Here $v=\beta c$ is the speed of two symmetric blobs, where $\theta$ is the viewing angle of the approaching blob. 
From the flux density of the hot spot components within models\,B to D, we obtain $R= 1.65^{+0.03}  _{-0.05}$.
For illustrative purposes, we initially consider a broad inclination angle between $0^\circ \leq \theta \leq 25^\circ $ and consider three representative cases for hot spot speeds, $\beta _{\rm HS}$= 0.053, 0.060, and 0.067, which are colour-coded in Fig.~\ref{fig:parameter_space}. 
The spectral index in the range obtained from Fig. \ref{fig:spixlofarvla} is encoded in the opaqueness of each colour, where a more saturated tone indicates a steeper $\alpha$ ($\alpha_{min}=-1.4$; $\alpha_{max}=-0.84$).
A lower limit of $\theta _{\mathrm{min}} \gtrsim 2^\circ$ is obtained by comparing the measured projected size of the source with the largest known radio galaxies \citep[ $\sim 1~$Mpc;][]{jeyakumar2000} by assuming that these represent an upper limit on the physical size a radio galaxy can have.
The SED of the core (see Sect.~\ref{sec:modeling}) then suggests that the inclination angle of the inner jet to be around $\theta _{\mathrm{SED}} \approx 3^\circ$. While the inclination angle can be deviated further along the jet due to various effects (e.g. jet bending or precession), we see that the estimates of hot spot speeds within this model depend only weakly on the orientation angle, and thus cannot deviate much from about 0.06\,$c$.
\\

\section{Jet age}
\label{app:age}
The evolutionary stage, and therefore the age, of a jet is an important characteristic. It can be estimated from its size, assuming that estimates of the hot spot advance speed derived from SED modelling are accurate and that this advance speed is constant in time. Despite the failures of models\,B to D, we note that their estimates of the hot spot advance speed should be correct to within an order of magnitude \citep{scheuer1995}.
The results of the following derivation depend linearly on speed; they too are expected to be correct to within an order of magnitude.
\citet{perucho2019} shows that numerical simulations of jet evolution in hot galactic atmospheres find that jets may accelerate down the pressure gradient of the galactic halo, and that they also undergo deceleration as the dentist drill effect develops \citep{scheuer1974}. The assumption of a constant velocity is thus known not to hold, but it can still provide a good first-order estimate. In this case, the age of the jet is given by
\begin{equation}
    t_{\rm jet} \simeq 7.3\times10^6 \left(\frac{L_{\rm jet}}{135\,{\rm kpc}}\right)\left(\frac{0.06}{\beta_{\rm HS}}\right)\,{\rm yr},
\end{equation}
where $L_{\rm jet}$ is the jet length. 

Since powerful relativistic jets inject a large part of their energy into lobe pressure \citep[see][]{perucho2017}, we can use the following expression, derived from dimensional arguments \citep[see][]{begelman1989,perucho2017}, to estimate the lobe pressure from the power and age of the jet as well as the volume of the lobe,
\begin{equation}
    p_{\rm lobe} = \kappa \frac{P_{\rm jet}\,t_{\rm jet}}{V_{\rm lobe}},
\end{equation}
with $\kappa\simeq 0.4$ for relativistic jets \citep[see][]{perucho2017}. Approximating the lobe volume to that of a cylinder of  length $L$ and an estimated lobe radius of $20\,{\rm kpc}$, taking $P_{\rm jet} \simeq 10^{47}\,{\rm erg/s}$, and using the previous expression for the age of the jet, we obtain 
\small
\begin{equation}
    p_{\rm lobe} \simeq 1.8\times10^{-9} \left(\frac{\kappa}{0.4}\right)\, \left(\frac{P_{\rm jet}}{10^{47}\,{\rm erg/s}}\right)\,\left(\frac{\beta_{\rm HS}}{0.06}\right)^{-1}\,\left(\frac{R_{\rm lobe}}{20\,{\rm kpc}}\right)^{-2}\,{\rm erg/cm^3}.
\end{equation}
\normalsize
The expression is independent of jet length because the contributions of both jet age and lobe volume cancel out. This estimate of lobe pressure is two orders of magnitude higher than that of the magnetic pressure implied by the modelled magnetic field strengths ($\simeq 2\times10^{-12}\,{\rm erg/cm^3}$). This could be explained by considering that the lobe pressure derived here would also include the contribution of the thermal population, which can dominate over both the electron and magnetic pressures \citep{croston2005}.  
\cite{hardcastle2015} provides evidence that the magnetic field should be around equipartition with the emitting particles (electrons) in hot spots and lobes of different FR-II sources, which is in apparent conflict with our result. Taking into account that the jet power cannot be significantly smaller than $P_{\rm jet}\simeq 10^{47}\,{\rm erg\,s^{-1}}$, and that the kinematic estimate of the jet age should be correct to an order of magnitude, we can see that both approaches can only be reconciled if the parameter $\kappa \ll 1$ and/or the total lobe pressure is significantly larger than that of the magnetic field (i.e. if the lobes are dominated by the thermal population). \citet{perucho2017} indicate that it is very unlikely that $\kappa \ll 1$, even for non-relativistic jets; the lobes could conceivably be dominated by the thermal population, however. Higher-quality data will be needed to properly verify this hypothesis.

\end{appendix}

\end{document}